\pdfoutput=1

\documentclass[11pt,twoside,a4paper,cmspaper,final,collab]{cms-tdr}
\def\svnVersion{747a49d-D}\def\svnDate{2019/08/11}\def\cmsCernNoTag{CERN-EP-2019-082}\def\cmsCernDate{\today}\def\cmsMessage{"Published in Physics Letters B as \href{http://dx.doi.org/10.1016/j.physletb.2019.134811}{\doi{10.1016/j.physletb.2019.134811}.}"}

\begin{document}\cmsNoteHeader{HIG-18-025}

\hyphenation{had-ron-i-za-tion}
\hyphenation{cal-or-i-me-ter}
\hyphenation{de-vices}
\RCS$HeadURL: svn+ssh://svn.cern.ch/reps/tdr2/papers/HIG-18-025/trunk/HIG-18-025.tex $
\RCS$Id: HIG-18-025.tex 494691 2019-05-20 18:36:14Z spanier $

\newcommand{\PgUn}{\ensuremath{\PgU}\mathrm{(nS)}}

\newlength\cmsFigWidth
\ifthenelse{\boolean{cms@external}}{\setlength\cmsFigWidth{0.85\columnwidth}}{\setlength\cmsFigWidth{0.4\textwidth}}
\ifthenelse{\boolean{cms@external}}{\providecommand{\cmsLeft}{upper\xspace}}{\providecommand{\cmsLeft}{left\xspace}}
\ifthenelse{\boolean{cms@external}}{\providecommand{\cmsRight}{lower\xspace}}{\providecommand{\cmsRight}{right\xspace}}

\cmsNoteHeader{HIG-18-025}
\title{Search for Higgs and $\cPZ$ boson decays to $\JPsi$ or $\PgU$ pairs in the four-muon final state in proton-proton collisions at $\sqrt{s}=13\TeV$}

\date{\today}
\abstract{A search for decays of the Higgs and $\cPZ$ boson to pairs of $\JPsi$ or $\PgUn$ (n=1,2,3) mesons, with their subsequent decay to \MM pairs, is presented. The analysis uses data from proton-proton collisions at $\sqrt{s}=13\TeV$, collected with the CMS detector at the LHC in 2017 and corresponding to an integrated luminosity of 37.5\fbinv. While an observation of such a decay with this sample would indicate the presence of physics beyond the standard model, no significant excess is observed. Upper limits at 95\% confidence level are placed on the branching fractions of these decays. In the $\JPsi$ pair channel, the limits are $1.8\times 10^{-3}$ and $2.2\times 10^{-6}$ for the Higgs and $\cPZ$ boson, respectively, while in the combined $\PgUn$ pair channel, the limits are $1.4\times 10^{-3}$ and $1.5\times 10^{-6}$, respectively, when the mesons from the Higgs and $\cPZ$ boson decay are assumed to be unpolarized. When fully longitudinal and transverse polarizations are considered the limits reduce by about 22--29\% and increase by about 10--13\%, respectively.}

\hypersetup{
pdfauthor={CMS Collaboration},
pdftitle={Search for Higgs and Z boson decays to J/Psi or Y pairs in proton-proton collisions at
sqrt(s) = 13 TeV},
pdfsubject={Rare decays},
pdfkeywords={CMS, standard model physics, Higgs boson, Z boson, rare decays}}

\maketitle

\section{Introduction}
\begin{figure*}[htp]
\begin{center}
\begin{tabular}{cccc}
    \includegraphics[width=0.227\textwidth]{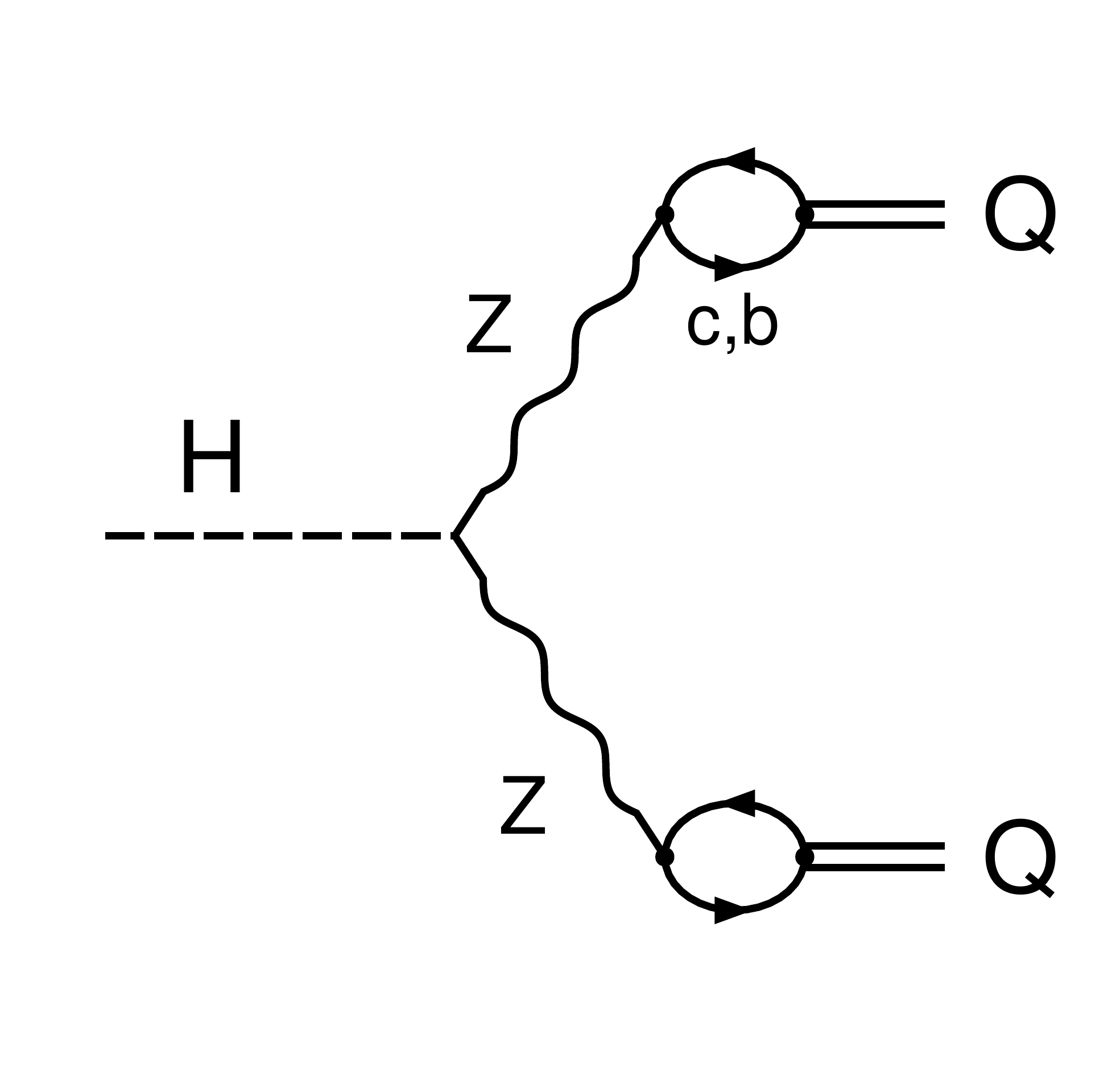} &
    \includegraphics[width=0.227\textwidth]{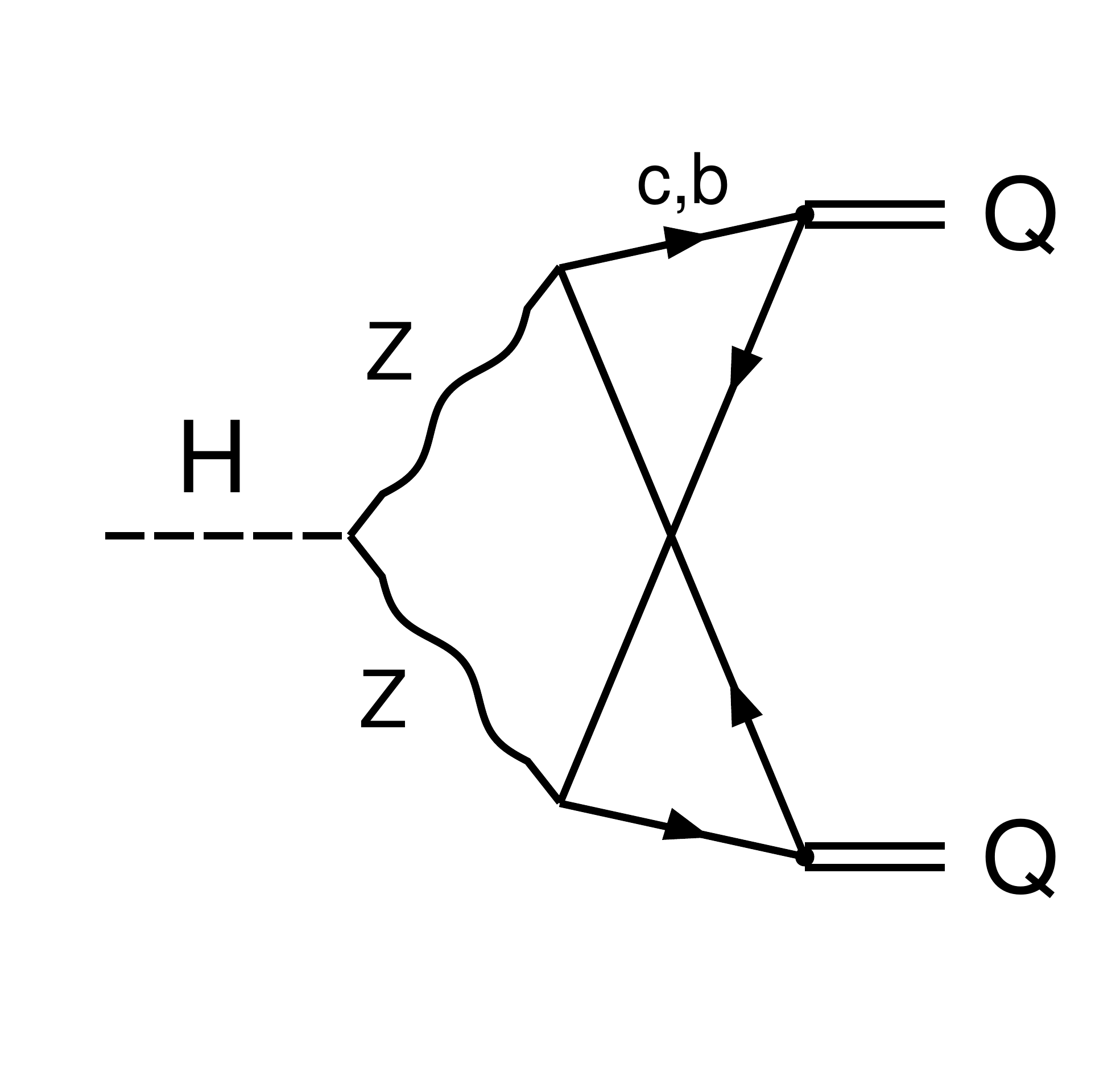} &
    \includegraphics[width=0.227\textwidth]{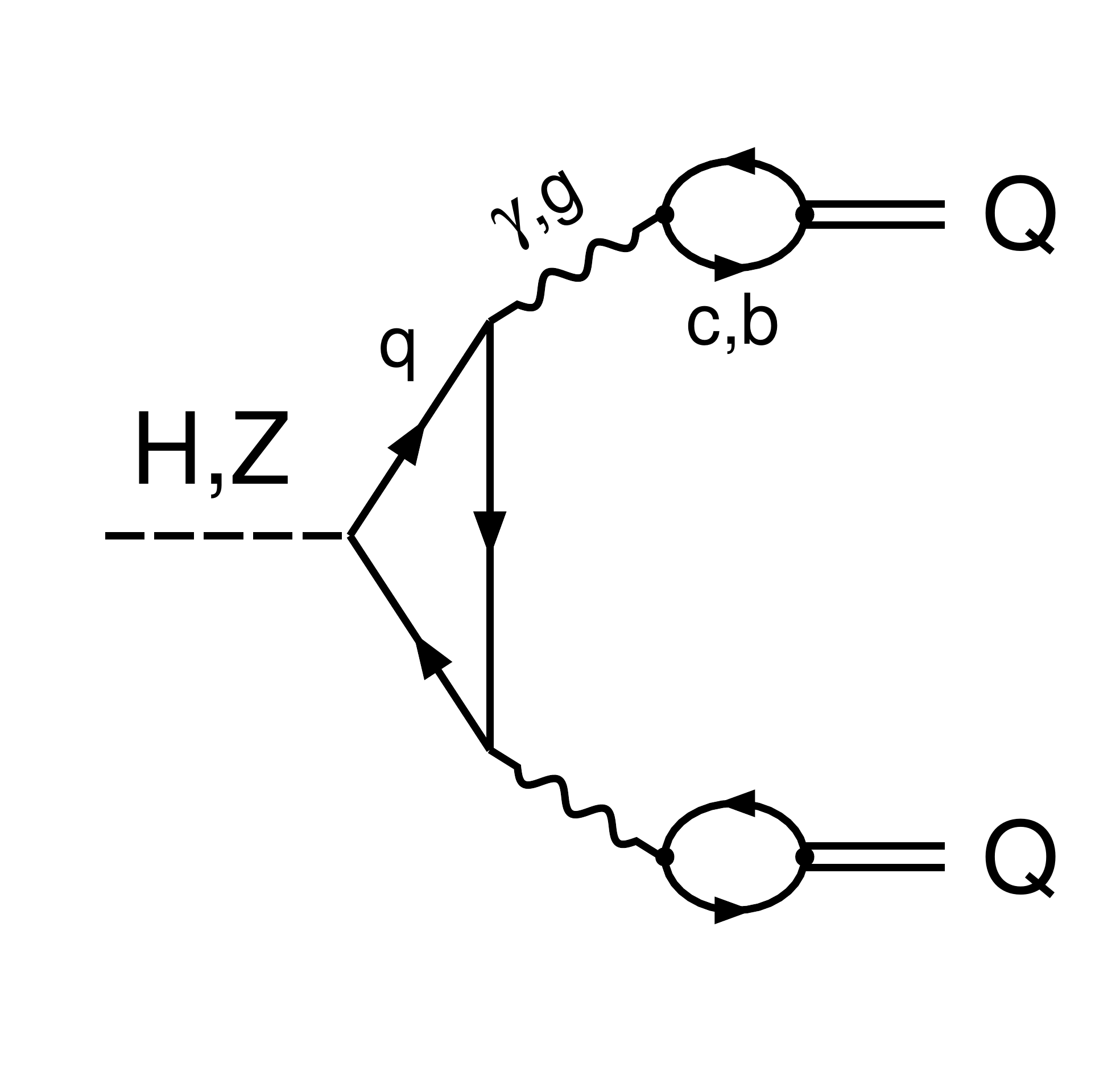} &
    \includegraphics[width=0.227\textwidth]{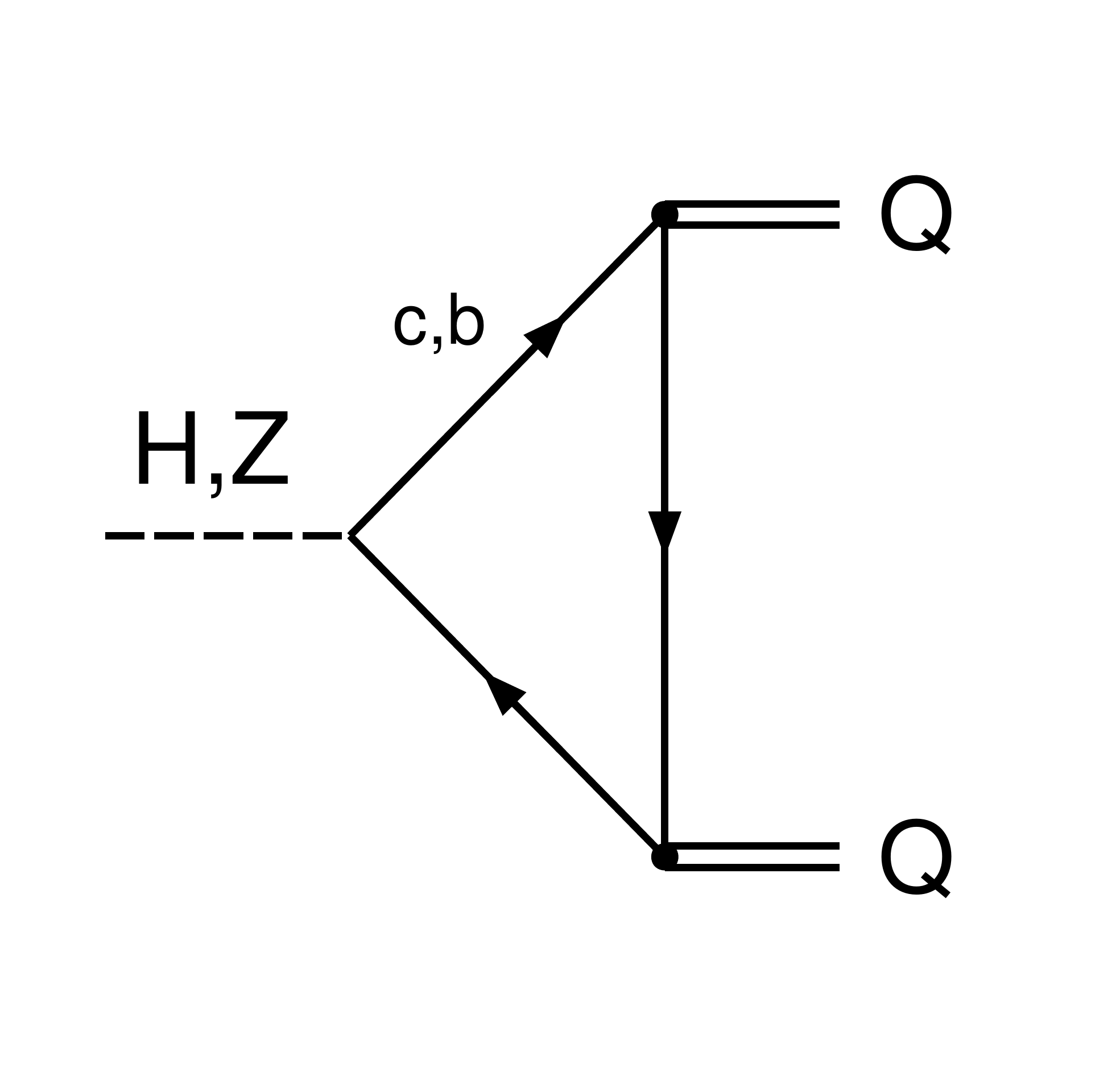} \\
\end{tabular}
 \caption{Feynman diagrams for $\PH\to\text{QQ}$ and $\cPZ\to\text{QQ}$
with Q = charmonium or bottomonium states.
In the two leftmost diagrams \cite{kartveli}, the virtual particles are 
$\cPZ$ bosons. The center-right diagram depicts indirect processes in the case where 
the flavor of the quark $\Pq$ in the loop is top, and direct processes in the case where
it is charm or bottom. The other virtual particles are either photons or gluons. In the
latter case additional soft-gluon exchange occurs. The rightmost diagram shows 
direct processes for the Higgs or $\cPZ$ bosons.
\label{fig:feynman}}
\end{center}
\end{figure*}

A new boson with a mass of 125\GeV was discovered by the ATLAS and CMS Collaborations at
the CERN LHC in 2012~\cite{Chatrchyan2013,Aad:2013xqa,201230,Chatrchyan:2013lba,CMS:2014ega,AtlasProperties,CMS:2015kwa}.
Comprehensive studies in various decay channels and production
modes have shown that the properties of the new boson are consistent, so far,
with expectations for the standard model (SM)
Higgs boson (\PH)~\cite{Aad:2015zhl,CMS:2015kwa,CMS:2018lkl}.
Recently, the Higgs boson couplings to top and bottom quarks
have been directly measured~\cite{HttCMS,HttATLAS,HbbCMS,HbbATLAS}.
Couplings to lighter quarks are still not observed directly.
Rare exclusive decays of the Higgs boson to mesons provide experimentally clean
final states to study Yukawa couplings to quarks and physics beyond the SM (BSM).
Examples of diagrams for decays of the Higgs and $\cPZ$ boson into quarkonium 
pairs are shown in Fig.~\ref{fig:feynman}.
The symbol Q refers to charmonium and bottomonium states.

The importance of the measurement of such decays has been
pointed out by Ref.~\cite{soni1979,keung1983,doroshenko1987,kartvili1988}.
Using a phenomenological approach for the direct \PH-$\qqbar$
coupling, Ref.~\cite{soni1979} finds that the dominant quarkonium
pair decay mode is $\PH\to \PgU \PgU$ with an estimated 
branching fraction ($\mathcal{B}$) at the level of $10^{-5}$.
The early calculations of Higgs boson decays into a pair of heavy
quarkonia states did not include relativistic corrections
caused by the internal motion of quarks~\cite{kartveli}.
The importance of the latter corrections is underlined by the fact that
the predicted $\re^+\re^- \to \JPsi \eta_c$ cross section increases by an order of
magnitude~\cite{BONDAR2005215,braguta2005,braguta2006} when these effects are included,
in agreement with measurements by the Belle and BaBar
experiments~\cite{belle2002,babar2005}.

With emphasis on amplitudes where the Higgs boson couples indirectly
to the final state me\-sons, such as represented by the two leftmost diagrams
in Fig.~\ref{fig:feynman}, Ref.~\cite{kartveli} arrives at values
of about $\mathcal{B}(\PH\to \JPsi \JPsi) = 1.5 \times 10^{-10}$ and
$\mathcal{B}(\PH\to \PgU \PgU) = 2 \times 10^{-9}$.
The mechanism where the Higgs boson couples directly to charm or bottom quarks, which then
hadronize to heavy quarkonia, was considered in a recent calculation~\cite{bodwin2013}
leading to an increase of an order of magnitude in $\mathcal{B}(\PH\to \JPsi \PGg)$.
The Higgs boson decay to the $\JPsi$ pair could also occur when the photon
in the $\JPsi \PGg$ decay is virtual and transforms into a $\JPsi$ meson.
Recently, the decay $\PH\to \JPsi \PGg$ has been searched for by the ATLAS and CMS 
collaborations~\cite{atlashjg,cmshjg}.
This Letter also presents the first search for decays of the $\cPZ$ boson into
quarkonium pairs. Feynman diagrams are shown in Fig.~\ref{fig:feynman}
(two rightmost plots). The SM prediction for $\mathcal{B}(\cPZ\to\JPsi \JPsi)$ 
calculated in the framework of nonrelativistic QCD and leading twist light cone
models is of the order of $10^{-12}$~\cite{zdecaytheory}. Several approximations 
for the non-perturbative QCD processes are used, including the restriction to 
color-singlet quarkonium states.

New physics could affect the direct boson couplings
or could enter through loops, and alter the interference
pattern between the amplitudes. Any of those possibilities enhance
branching fractions with respect to the SM predictions.
Many BSM theories predict substantial modifications of the Yukawa
couplings of the Higgs boson to quarks,
such as models with Higgs-dependent Yukawa couplings~\cite{giudice}, the
minimal flavor violation framework~\cite{ambrosio},
the Froggatt--Nielsen mechanism~\cite{froggatt},
and the Randall--Sundrum family of models~\cite{randall}.
An overview of models can be found in Ref.~\cite{LHC-YR4}.
In the related quarkonium--$\PGg$ channels, deviations of the
\PH-$\qqbar$ couplings from the SM predictions can change
the interference between direct and indirect amplitudes,
resulting in substantial modifications of the branching fractions, particularly in the
$\PgU$ channel, where the increase is up to several orders of
magnitude~\cite{bodwin2013}.
The observation of a Higgs or $\cPZ$ boson signal in the quarkonium pair decay
modes with the available LHC data sets would indicate the presence of BSM physics.

This Letter presents the first search for the Higgs and $\cPZ$ boson
decays into $\JPsi$ or $\PgU$ meson pairs, where $\PgU$ stands for
the combined contribution of the $\PgUn$ states with n = 1,2,3.
The subsequent decay of these meson pairs to the 4\Pgm\
final state offers a very clean experimental signature that is used in this analysis.
For the  $\JPsi$  meson pairs, feed-down from higher charmonium
states are not taken into account.  For the $\PgUn$ meson pairs, decays from higher
to lower mass $\PgUn$ states are included. The results presented in this Letter are
based on proton-proton ($\Pp\Pp$) collision data recorded in 2017 with the CMS
detector at a center-of-mass energy of $\sqrt{s}=13\TeV$, amounting to an
integrated luminosity of 37.5\fbinv.

\section{The CMS detector}
A detailed description of the CMS detector, together with a definition
of the coordinate system used and the relevant kinematic variables,
can be found in Ref.~\cite{CMS-Jinst}. The central feature of the CMS
apparatus is a superconducting solenoid, 13\unit{m} in length and
6\unit{m} in internal diameter, providing an axial magnetic field of
3.8\unit{T}. Within the solenoid volume are a silicon pixel and strip
tracker, a lead tungstate crystal electromagnetic calorimeter,
and a brass and scintillator hadron calorimeter, each composed
of a barrel and two endcap sections. Forward calorimeters extend the
pseudorapidity ($\eta$) coverage provided by the barrel and endcap
detectors. Muons are detected in gas-ionization chambers embedded in
the steel flux-return yoke outside the solenoid.
They are measured in the range $\abs{\eta} < 2.4$, with
detection planes made using three technologies: drift tubes, cathode strip
chambers, and resistive-plate chambers.

An entirely new pixel detector has been installed after 2016, featuring
an all-silicon device with four layers in the barrel and three disks in the
endcaps~\cite{newpixel}, providing four pixel detector measurements.
Reduced material budget in front of the calorimeters was achieved with 
two-phase CO$_2$ cooling and light-weight mechanical support, 
and moving the electronic boards and connections out of the tracking volume.

Events of interest are selected using a two-tiered trigger
system~\cite{Khachatryan:2016bia}. The first level, composed
of custom hardware processors, uses information from the calorimeters
and muon detectors to select events at a rate of around 100\unit{kHz}
within a time interval of less than 4\mus. The second level, known as
the high-level trigger, consists of a farm of processors running a version
of the full event reconstruction software optimized for fast processing,
and reduces the event rate to around 1\unit{kHz} before data storage.

Dedicated triggers were deployed in 2017 to enhance the selection of events
of interest for the present study. They require the presence of at least
three muons with \pt greater than 2\GeV. Two of these must be oppositely charged
and have to originate from a common vertex with a probability greater than 0.5\%,
as determined by a Kalman vertex fit~\cite{kalman}, thus suppressing random combinations 
of two muons. The $\JPsi$-specific trigger requires a dimuon system's invariant mass
to be between 2.95 and 3.25\GeV and its \pt to be greater than 3.5\GeV.
The trigger used to select the $\PgU$ sample requires
two of the three muons to have \pt
greater than 3.5\GeV, and one muon \pt greater than 5\GeV.
The invariant mass for one oppositely charged muon pair must lie in the
interval 8.5--11.4\GeV.
Both triggers gave an efficiency exceeding 85\% to select events satisfying
the selection criteria used in the analysis.

\section{Signal and background modeling}
Simulated samples of the Higgs and $\cPZ$ boson signals are used to estimate the
expected signal yields and model the distribution of signal events in the four-muon
invariant mass. For the $\PH\to\JPsi \JPsi$ and $\PH\to\PgU \PgU$ samples
the Higgs boson is produced with the \POWHEG v2.0 Monte Carlo (MC) event
generator~\cite{Alioli:2008tz,Nason:2009ai}, which includes
the gluon-gluon fusion ($\Pg\Pg$F) and vector-boson fusion production processes.
The parton distribution function (PDF) set used is NNPDF3.1~\cite{Ball:2017nwa}.
The JHUGen 7.1.4 generator~\cite{jhugenerator1,jhugenerator2} is used to decay the
Higgs boson into two vector mesons taking into account their helicity.
To produce the decay for unpolarized quarkonia, the JHUGen generator
is configured to model a uniform muon helicity angle distribution.
The generator is interfaced with \PYTHIA 8.226~\cite{Sjostrand:2014zea}
for parton-showering and hadronization according to the CUETP8M1~\cite{Khachatryan:2015pea} tune.
The total SM Higgs boson production cross section for the calculation of
branching fractions is taken from the LHC Higgs cross section working group~\cite{LHC-YR4}.

The $\cPZ\to\JPsi \JPsi$ and $\cPZ\to\PgU \PgU$
samples are produced with the \PYTHIA 8.226 generator~\cite{Sjostrand:2014zea}, tune
CUETP8M1~\cite{Khachatryan:2015pea}.
The SM $\cPZ$ boson production cross section includes the
next-to-next-to-leading order (NNLO) QCD contributions, and
the next-to-leading order (NLO) electroweak
corrections from \FEWZ 3.1~\cite{fewz} calculated using the NLO PDF set NNPDF3.0.
The \cPZ\ boson \pt is reweighted to match the NLO
calculation~\cite{Alioli:2008tz,Nason:2009ai,Alwall:2014hca}.
The total cross section is obtained with the $\mathcal{B}(\cPZ\to\MM )$ value
from Ref.~\cite{pdg}.

In the $\JPsi$ and \PgU\ pair channels backgrounds are assumed to originate
from prompt nonresonant pair production, which in $\Pp\Pp$ collisions
dominantly occurs via $\Pg\Pg$F~\cite{lansberg,Baranov:2013,qiao,cmsjj}.
Initially, the two mesons are color-octet bound states that then
radiate soft gluons to become real mesons. Event samples are generated
according to this model~\cite{qiao}.

The generated events are processed through a detailed simulation of the CMS detector based
on \GEANTfour~\cite{GEANT4}. The high instantaneous luminosity of the LHC results in
multiple $\Pp\Pp$ interactions per bunch crossing.
Simultaneous $\Pp\Pp$ interactions that overlap with the event of interest, \ie pileup, are
included in simulated samples. The distribution of the number of additional
interactions per event in simulation corresponds to that observed in the data,
where the average pileup number is found to be 32. 

The acceptance of the final states changes with the angular distribution of
the muons in the quarkonium decay.
The distribution of the decay angle $\theta$, defined as the angle between
the positive muon direction of flight in the rest frame of the quarkonium
with respect to the quarkonium direction in the boson rest frame, is proportional
to $(1 + \lambda_\theta \cos^2\theta)$.
In this Letter, the nominal results are obtained using a signal acceptance
calculated for the unpolarized case $(\lambda_\theta = 0)$. Two extreme scenarios
have also been considered, where the $\JPsi$ and $\PgU$ mesons are either fully
transversely polarized, $\lambda_\theta = +1$, or fully longitudinally polarized,
$\lambda_\theta = -1$. No azimuthal anisotropies have been considered.
According to Refs.~\cite{kartveli,zdecaytheory} the $\JPsi$ and $\PgU$
mesons produced in the decays of both bosons are expected to be dominantly
longitudinally polarized.

\section{Data reconstruction and selection}
Muons are reconstructed by combining information from the silicon tracker and
the muon system~\cite{Sirunyan:2018fpa}.
The matching between tracks reconstructed in each of the subsystems proceeds
either outside-in, starting from a track in the muon system, or inside-out,
starting from a track provided by the silicon tracker.
In the latter case, tracks that match track segments in
only one or two stations of the muon system are also considered in the analysis to
collect very low-\pt muons that may not have sufficient energy to penetrate the
entire muon system. The muons are selected from the reconstructed muon track
candidates that match with at least one segment in any muon station in both $x$ and $y$.
The number of silicon tracker layers with hits used in the muon track candidate
has to be greater than 5 and include at least one pixel detector layer.
Matching muons to tracks measured in the silicon tracker results in a relative
transverse momentum (\pt) resolution of 1\% in the barrel and 3\% in the endcaps
for muons with \pt up to 100\GeV. The \pt resolution in the barrel is better
than 7\% for muons with \pt up to 1\TeV~\cite{Sirunyan:2018fpa}.

The reconstructed vertex with the largest value of summed
charged particle $\pt^2$ is taken to be the primary $\Pp\Pp$ interaction vertex.
To suppress muons originating from nonprompt hadron decays,
the impact parameter of each muon track, computed with respect to the position of
the primary $\Pp\Pp$ interaction vertex, is required to be less than 0.3 (20.0)\unit{cm} in
the transverse plane (longitudinal axis).
Events with at least four such muons with $\pt>3\GeV$ and $\lvert\eta\rvert <2.4$ are
accepted. To isolate the leading muon candidate from other hadronic activity
in the event, a cone of size ${\Delta}R=\sqrt{\smash[b]{(\Delta\eta)^{2}+(\Delta\phi)^{2}}} = 0.3$ is constructed around its momentum direction, where $\phi$ is the azimuthal angle in radians.
The sum of the \pt of the reconstructed inner-detector tracks originating from the
primary $\Pp\Pp$ interaction vertex within the cone has to be less than 50\% of the muon's \pt.
The transverse momentum of the leading muon is subtracted from the sum
and the subleading muon \pt is also subtracted, if this muon falls within
the isolation cone of the leading muon.

\begin{figure*}[h]
  \begin{center}
    \includegraphics[width=0.49\textwidth]{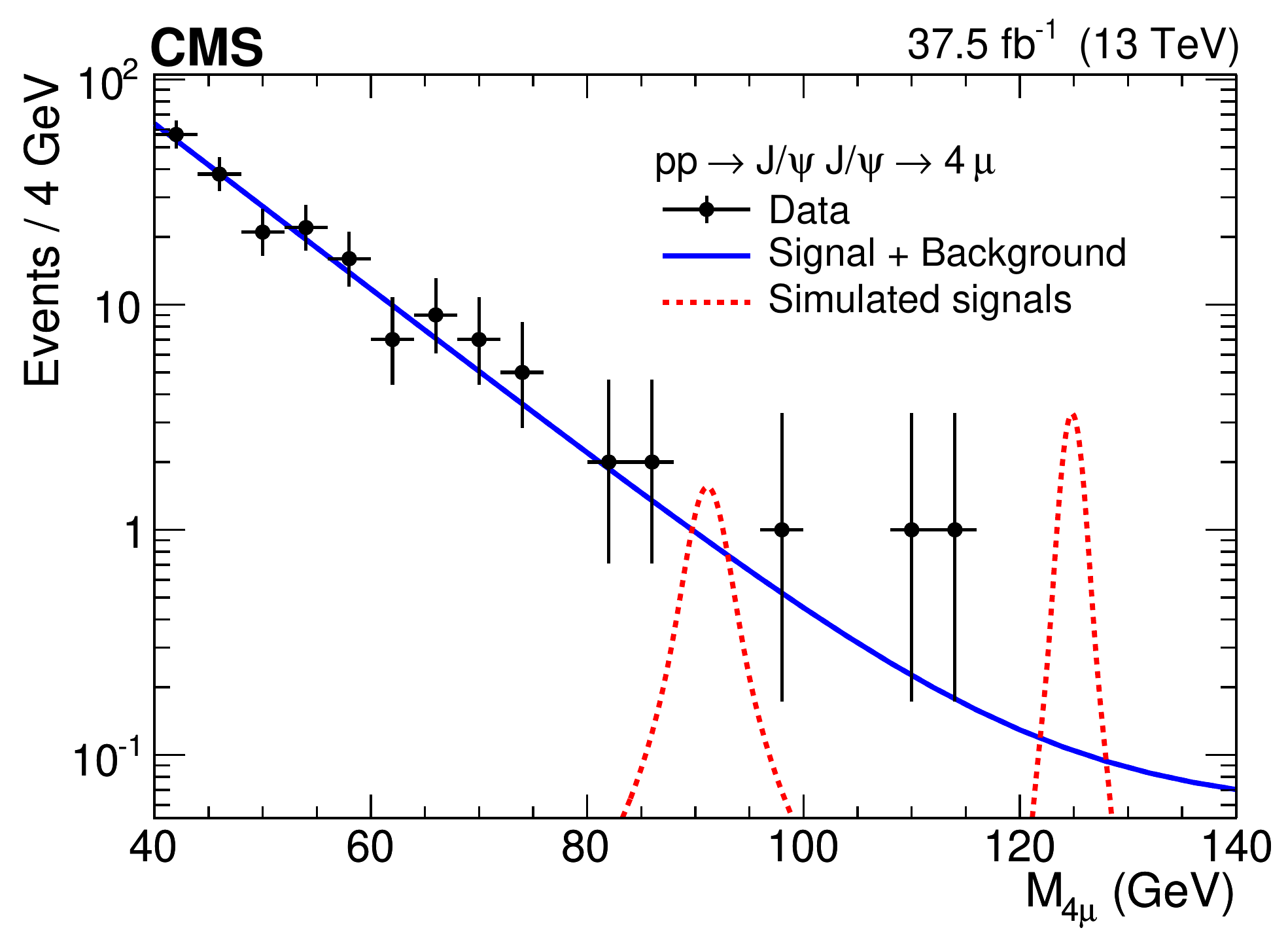}
    \includegraphics[width=0.49\textwidth]{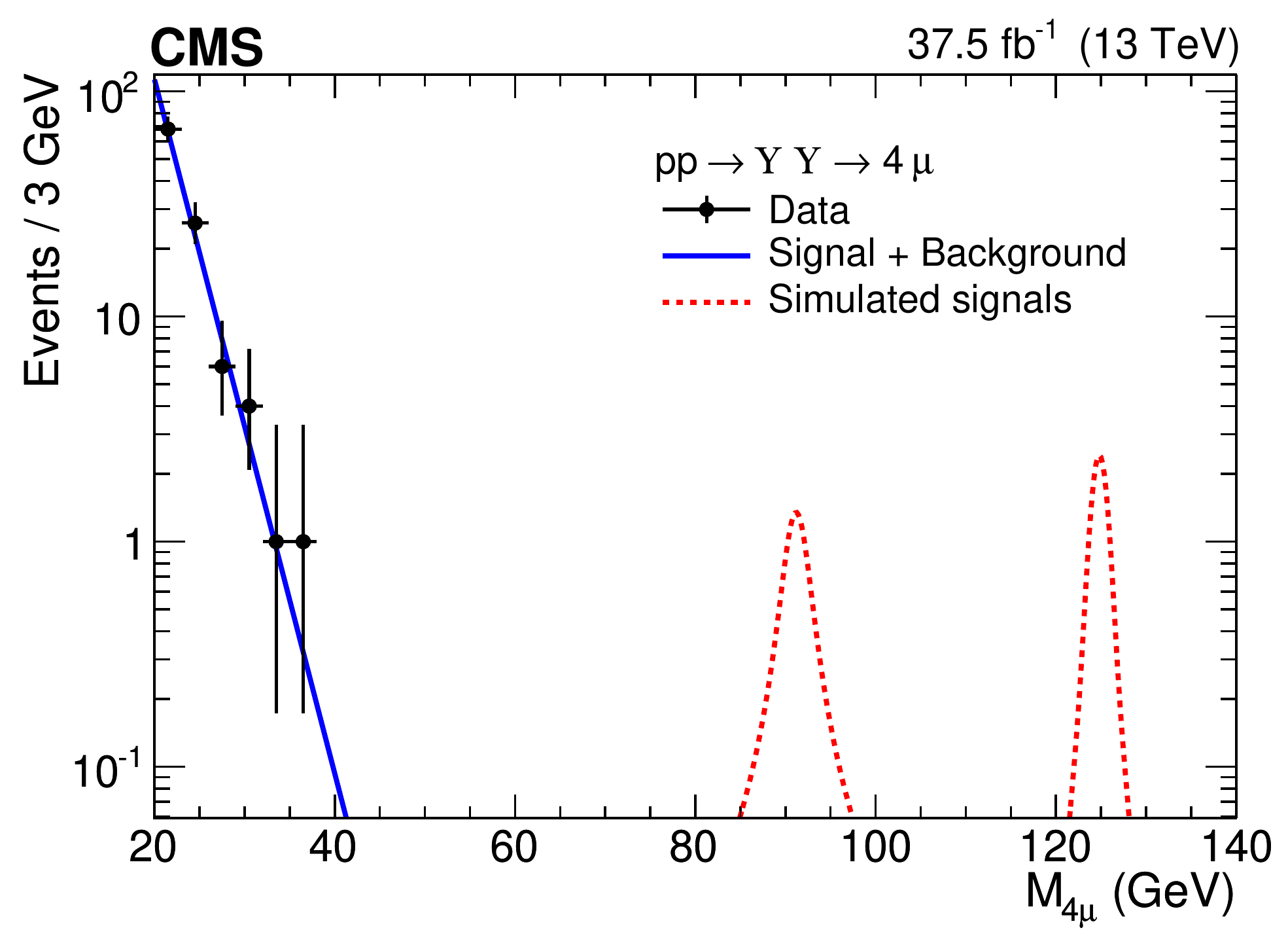}
    \caption{The four-muon invariant mass distributions, for $\JPsi \JPsi$ (left) and
$\PgU \PgU$ (right) candidates (error bars for zero entries are omitted).
The result of the maximum likelihood fit is superimposed
(solid blue line). For illustrative purposes, the plots also show the distributions
for simulated Higgs and $\cPZ$ boson signals (dashed red lines), each normalized
to two events. \label{fig:kineplot1}
}
  \end{center}
\end{figure*}

The $\JPsi$ and $\PgU$ candidates are built from pairs of oppositely charged muons.
Each muon pair must fit to their common vertex with a probability greater than 0.5\%.
The $\JPsi$ candidate's \pt has to be greater than 3.5\GeV, matching the trigger requirement,
and the invariant masses of the higher and lower-\pt $\JPsi$ candidates have to be within
0.1 and 0.15\GeV, respectively, of the nominal mass of the $\JPsi$. The dimuon mass resolution
is about 30\MeV. The mass window of the subleading $\JPsi$ is wider to allow further
monitoring of the sideband population.
To suppress contributions from nonprompt hadrons, separately produced $\JPsi$s
and muons from other sources, the four-muon Kalman vertex fit probability of $\JPsi$
pairs has to be greater than 5\%.
Finally, the absolute value of the difference in rapidity between the two $\JPsi$
candidates has to be less than 3. This criterion marginally affects the
signal while removing about 20\% of  the selected events.
After the selection, 189 events are found in data in the 40--140\GeV four-muon invariant
mass range. Figure~\ref{fig:kineplot1} (left) shows the four-muon invariant mass distribution.

For the selection of $\PgU$ pair candidates, the same event selection criteria are
applied, except that the $\PgU$ candidate \pt has to be greater than 5\GeV,
and the invariant mass has to fall within the range 8.5--11\GeV.
Furthermore, the four-muon Kalman vertex fit probability has to be greater than
1\% to suppress random combinations. The nonprompt background is
negligible in this channel. After applying the selections, 106 events are found in
data in the 20--140\GeV four-muons invariant mass range.
Figure~\ref{fig:kineplot1} (right) shows the four-muon invariant mass distribution.

The differences in efficiencies between data and simulation for the trigger,
offline muon reconstruction, identification, and isolation are corrected by
reweighting the simulated events with data-to-simulation correction factors,
which are obtained with the ``tag-and-probe'' method~\cite{cite:tagandprobe}
using $\JPsi\to\MM$ events. The scale correction factors are observed to
deviate from unity by less than 3\%. The difference in the four-muon Kalman vertex
fit efficiency between data and simulation is evaluated with $\JPsi$ pair
event samples and found to be less than 3\%.
The total signal efficiency, including kinematic
acceptance, trigger, reconstruction, identification, and
isolation efficiencies, for the $\JPsi \JPsi$ decays
with unpolarized $\JPsi$ is approximately 23\% for both bosons.
For the $\PgU \PgU$ decays the corresponding efficiency is about 27\%.

\section{Results}
Unbinned extended maximum-likelihood fits~\cite{roofit} to the four-muon invariant mass distributions $M_{4\Pgm}$ are performed. Yields for signals and backgrounds are free parameters in the fit.
For the Higgs boson the invariant mass distribution obtained from simulation is described with two
Gaussian functions with a common mean. The simulated $\cPZ$ signal is described with a Voigtian
function with the world-average value for the resonance width~\cite{pdg}.
The mass resolution and mean are taken from the fit to the simulation,
and they are fixed in the fit to data.

The four-muon invariant mass distribution up to 140\GeV is described
by an exponential plus constant function. The relative contribution and
decay constant of the exponential function are varied in the fit to data.
The values of both parameters are found to be in close agreement between
observation and simulation~\cite{qiao}.
The result of the fit is shown as a solid blue line in Fig.~\ref{fig:kineplot1} (left).

In the $\PgU$ pair sample, no events are observed above
the four-muon invariant mass of 40\GeV. The four-muon invariant mass distribution
is modeled analogously to the $\JPsi$ pair channel. The $M_{4\Pgm}$ distribution below
40\GeV is well described by an exponential function.
The decay constant of the exponential function is also varied in the fit.
The same function describes an event sample generated with the pair production model~\cite{qiao}.
Figure~\ref{fig:kineplot1} (right) shows the observed $M_{4\Pgm}$ distribution
with the fit superimposed.

Given the absence of a signal for either of the bosons, upper
limits on the branching fractions are obtained.
They are set by using the modified frequentist approach,
\CLs, with the profile likelihood ratio as a test statistic~\cite{junk,read,combiner}.
The uncertainties affecting the signal yields include the contributions
from the luminosity measurement~\cite{CMS-PAS-LUM-17-004},
the corrections applied to the simulated events in order to compensate for
differences in trigger, muon reconstruction and identification efficiencies,
momentum scale and resolution of muon candidates, and
four-muon vertex fit.
Sources for theoretical uncertainties are the QCD coupling
and PDF choice~\cite{LHC-YR4,Ball:2017nwa,Butterworth:2015oua},
and the renormalization and factorization scale
choice~\cite{Butterworth:2015oua,Martin2009,PhysRevD.82.074024,BALL2011296}.
The uncertainties in the $\JPsi$ and $\PgU$ branching fractions to muon pairs
are taken from Ref.~\cite{pdg}.
The relative impact of the systematic uncertainties on the upper
limits is less than 2\% in all channels.

The value for $\mathcal{B}(\JPsi \to \MM)$ is taken from Ref.~\cite{pdg}.
This analysis does not distinguish between the three $\PgUn$ states.
To calculate their contribution to the corresponding $\PH$ and $\cPZ$ boson
branching fraction the coupling strength of the bosons to any $\PgUn$
pairing is assumed to be the same.
All $\PgU$ states can directly decay into muon pairs with the different branching fractions
taken from Ref.~\cite{pdg}. In addition, it is assumed that one of the $\PgU$ states could
be the result of a transition $\PgUc\to \PgUb$, $\PgUc\to \PgUa$, or $\PgUb\to\PgUa$
before decaying into muons~\cite{pdg}.

The observed and median expected exclusion limits for the
branching fractions at 95\% confidence level (\CL) for the $\PH$
and $\cPZ$ boson decays are listed in Table~\ref{results}.

\begin{table}[htp]
\begin{center}
\topcaption{\label{results}Exclusion limits at 95\% \CL for the branching fractions of the
$\PH$ and $\cPZ$ boson decays to $\JPsi$ or $\PgU$ mesons pairs. The second column lists
the observed limits. The third column shows the median expected limits with the upper
and lower bounds in the expected 68\% \CL intervals.}
\renewcommand{\arraystretch}{1.3}
\begin{tabular}{lcr}
\hline
 Process & Observed & \multicolumn{1}{c}{Expected} \\
\hline
$\mathcal{B}(\PH\to\JPsi \JPsi)$ & $1.8 \times 10^{-3}$ & $(1.8^{+0.2}_{-0.1}) \times 10^{-3}$ \\
$\mathcal{B}(\PH\to\PgU \PgU)$  & $1.4 \times 10^{-3}$ & $(1.4\pm 0.1) \times 10^{-3}$ \\
$\mathcal{B}(\cPZ\to\JPsi \JPsi)$ & $2.2 \times 10^{-6}$ & $(2.8^{+1.2}_{-0.7}) \times 10^{-6}$ \\
$\mathcal{B}(\cPZ\to\PgU \PgU)$ & $1.5 \times 10^{-6}$ & $(1.5\pm 0.1) \times 10^{-6}$ \\
\hline
\end{tabular}
\end{center}
\end{table}

The relative changes in the upper limits on the Higgs boson decay branching fractions
with respect to the case of unpolarized decay mesons
are about $-$22\% for fully longitudinally polarized $\JPsi$ and $\PgU$ mesons,
and $+$10\% for fully transversely polarized mesons.
For the $\cPZ$ boson the relative changes are about $-$29 ($-$26)\% for fully
longitudinally polarized $\JPsi$ ($\PgU$) mesons
and $+$13 ($+$12)\% for fully transversely polarized mesons.

\section{Summary}
In summary, this Letter presents the first search for decays of the
Higgs and $\cPZ$ boson to pairs of $\JPsi$ or $\PgUn$ (n=1,2,3) mesons,
with their subsequent decay to $\MM$ pairs.
Data from $\Pp\Pp$ collisions at $\sqrt{s}=13\TeV$, corresponding
to an integrated luminosity of 37.5\fbinv are used.
No excess has been observed above a small background in
the $\JPsi$ pair and with vanishingly small background in the $\PgU$
pair channels. The observed upper limits at 95\% confidence level
on the branching fractions for the Higgs boson decays
for unpolarized mesons are
$\mathcal{B}(\PH\to\JPsi \JPsi) < 1.8\times 10^{-3}$
and
$\mathcal{B}(\PH\to\PgU \PgU) < 1.4\times 10^{-3}$.
The observed upper limits
on the branching fractions for the $\cPZ$ boson decay
in the unpolarized scenario are
$\mathcal{B}(\cPZ\to\JPsi \JPsi) < 2.2\times 10^{-6}$
and
$\mathcal{B}(\cPZ\to\PgU \PgU) < 1.5\times 10^{-6}$,
where all three $\PgUn$ states are considered.
Extreme polarization scenarios give rise to variations
in the observed boson decay branching fractions between
$-$(22--29)\% for fully longitudinally polarized $\JPsi$
and $\PgU$ mesons and $+$(10--13)\% for fully
transversely polarized mesons.
This analysis is expected to motivate renewed calculations 
of the Higgs boson branching fractions for rare standard model 
decays, as only a few positive signal events would indicate 
the presence of physics beyond the standard model.

\begin{acknowledgments}
We congratulate our colleagues in the CERN accelerator departments for the excellent performance of the LHC and thank the technical and administrative staffs at CERN and at other CMS institutes for their contributions to the success of the CMS effort. In addition, we gratefully acknowledge the computing centers and personnel of the Worldwide LHC Computing Grid for delivering so effectively the computing infrastructure essential to our analyses. Finally, we acknowledge the enduring support for the construction and operation of the LHC and the CMS detector provided by the following funding agencies: BMBWF and FWF (Austria); FNRS and FWO (Belgium); CNPq, CAPES, FAPERJ, FAPERGS, and FAPESP (Brazil); MES (Bulgaria); CERN; CAS, MoST, and NSFC (China); COLCIENCIAS (Colombia); MSES and CSF (Croatia); RPF (Cyprus); SENESCYT (Ecuador); MoER, ERC IUT, PUT and ERDF (Estonia); Academy of Finland, MEC, and HIP (Finland); CEA and CNRS/IN2P3 (France); BMBF, DFG, and HGF (Germany); GSRT (Greece); NKFIA (Hungary); DAE and DST (India); IPM (Iran); SFI (Ireland); INFN (Italy); MSIP and NRF (Republic of Korea); MES (Latvia); LAS (Lithuania); MOE and UM (Malaysia); BUAP, CINVESTAV, CONACYT, LNS, SEP, and UASLP-FAI (Mexico); MOS (Montenegro); MBIE (New Zealand); PAEC (Pakistan); MSHE and NSC (Poland); FCT (Portugal); JINR (Dubna); MON, RosAtom, RAS, RFBR, and NRC KI (Russia); MESTD (Serbia); SEIDI, CPAN, PCTI, and FEDER (Spain); MOSTR (Sri Lanka); Swiss Funding Agencies (Switzerland); MST (Taipei); ThEPCenter, IPST, STAR, and NSTDA (Thailand); TUBITAK and TAEK (Turkey); NASU and SFFR (Ukraine); STFC (United Kingdom); DOE and NSF (USA). 

\hyphenation{Rachada-pisek} Individuals have received support from the Marie-Curie program and the European Research Council and Horizon 2020 Grant, contract Nos.\ 675440, 752730, and 765710 (European Union); the Leventis Foundation; the A.P.\ Sloan Foundation; the Alexander von Humboldt Foundation; the Belgian Federal Science Policy Office; the Fonds pour la Formation \`a la Recherche dans l'Industrie et dans l'Agriculture (FRIA-Belgium); the Agentschap voor Innovatie door Wetenschap en Technologie (IWT-Belgium); the F.R.S.-FNRS and FWO (Belgium) under the ``Excellence of Science -- EOS" -- be.h project n.\ 30820817; the Beijing Municipal Science \& Technology Commission, No. Z181100004218003; the Ministry of Education, Youth and Sports (MEYS) of the Czech Republic; the Lend\"ulet (``Momentum") Program and the J\'anos Bolyai Research Scholarship of the Hungarian Academy of Sciences, the New National Excellence Program \'UNKP, the NKFIA research grants 123842, 123959, 124845, 124850, 125105, 128713, 128786, and 129058 (Hungary); the Council of Science and Industrial Research, India; the HOMING PLUS program of the Foundation for Polish Science, cofinanced from European Union, Regional Development Fund, the Mobility Plus program of the Ministry of Science and Higher Education, the National Science Center (Poland), contracts Harmonia 2014/14/M/ST2/00428, Opus 2014/13/B/ST2/02543, 2014/15/B/ST2/03998, and 2015/19/B/ST2/02861, Sonata-bis 2012/07/E/ST2/01406; the National Priorities Research Program by Qatar National Research Fund; the Programa Estatal de Fomento de la Investigaci{\'o}n Cient{\'i}fica y T{\'e}cnica de Excelencia Mar\'{\i}a de Maeztu, grant MDM-2015-0509 and the Programa Severo Ochoa del Principado de Asturias; the Thalis and Aristeia programs cofinanced by EU-ESF and the Greek NSRF; the Rachadapisek Sompot Fund for Postdoctoral Fellowship, Chulalongkorn University and the Chulalongkorn Academic into Its 2nd Century Project Advancement Project (Thailand); the Welch Foundation, contract C-1845; and the Weston Havens Foundation (USA).
\end{acknowledgments}

\bibliography{auto_generated}

\cleardoublepage \appendix\section{The CMS Collaboration \label{app:collab}}\begin{sloppypar}\hyphenpenalty=5000\widowpenalty=500\clubpenalty=5000\vskip\cmsinstskip
\textbf{Yerevan Physics Institute, Yerevan, Armenia}\\*[0pt]
A.M.~Sirunyan$^{\textrm{\dag}}$, A.~Tumasyan
\vskip\cmsinstskip
\textbf{Institut für Hochenergiephysik, Wien, Austria}\\*[0pt]
W.~Adam, F.~Ambrogi, T.~Bergauer, J.~Brandstetter, M.~Dragicevic, J.~Erö, A.~Escalante~Del~Valle, M.~Flechl, R.~Frühwirth\cmsAuthorMark{1}, M.~Jeitler\cmsAuthorMark{1}, N.~Krammer, I.~Krätschmer, D.~Liko, T.~Madlener, I.~Mikulec, N.~Rad, J.~Schieck\cmsAuthorMark{1}, R.~Schöfbeck, M.~Spanring, D.~Spitzbart, W.~Waltenberger, C.-E.~Wulz\cmsAuthorMark{1}, M.~Zarucki
% ...
\vskip\cmsinstskip
\dag: Deceased\\
1:  Also at Vienna University of Technology, Vienna, Austria\\
% ...
\end{sloppypar}
\end{document}